\documentclass[a4paper,12pt]{article}
\usepackage{bm,subeqnarray,eqsection,indent,cite}

\usepackage{hyperref}
\usepackage{amsmath,amssymb,amsfonts}

\newif\ifpdf
\ifx\pdfoutput\undefined
\pdffalse 
\else
\pdfoutput=1 
\pdftrue
\fi

\ifpdf
\usepackage[pdftex]{graphicx}
\usepackage{epstopdf}
\else
\usepackage{graphicx}
\fi

\newcommand{\be}{\begin{equation}}
\newcommand{\ee}{\end{equation}}

\newcommand{\ef}[1]{\, #1}

\newcommand{\OSP}{\mathfrak{osp}}

\newcommand{\reff}[1]{(\ref{#1})}

\def\psibar{{\bar{\psi}}}

\def\bt{{\bf t}}

\def\bw{{\bf w}}

\def\ve{{\bf e}}
\def\bone{{\mathbf{1}}}

\newtheorem{defin}{Definition}[section]

\newtheorem{proposition}[defin]{Proposition}

\newtheorem{lemma}[defin]{Lemma}

\newtheorem{corollary}[defin]{Corollary}

\def\proof{\par\medskip\noindent{\sc Proof.\ }}

\newcommand{\qed}{\quad $\Box$ \medskip \medskip}

\def\R{{\mathbb R}}
\def\C{{\mathbb C}}

\newcommand{\scrf}{{\mathcal{F}}}

\newcommand{\scrt}{{\mathcal{T}}}

\newcommand{\He}{\hbox{He}}

\newenvironment{scarray}{
          \textfont0=\scriptfont0
          \scriptfont0=\scriptscriptfont0
          \textfont1=\scriptfont1
          \scriptfont1=\scriptscriptfont1
          \textfont2=\scriptfont2
          \scriptfont2=\scriptscriptfont2
          \textfont3=\scriptfont3
          \scriptfont3=\scriptscriptfont3
        
        \begin{array}{c}}{\end{array}}

\begin{document}

\title{%
Hyperforests on the Complete Hypergraph \\ by Grassmann Integral Representation}


\author{Andrea Bedini, Sergio Caracciolo and 
        Andrea Sportiello \\
{\scriptsize 
Dipartimento di Fisica dell'Universit\`a degli Studi di Milano
        and INFN, Sezione di Milano,}  \\[-2mm]
{\scriptsize via Celoria 16, I-20133 Milano, Italy}\\
{\tt \scriptsize
Andr{}ea.Bedini@mi.in{}fn.it,
S{}erg{}io.Cara{\phantom{\rule{0pt}{0pt}}}cci{}olo@mi.i{}nfn.it, 
Andr{}ea.Sportie{\phantom{\rule{0pt}{0pt}}}llo@mi.in{}fn.it}}

\bigskip
\medskip

\date{\today}

\maketitle

\begin{abstract}
We study the generating function  of rooted and unrooted hyperforests in a general complete hypergraph with $n$ vertices  by using a novel Grassmann representation of their generating functions.
We show that this new approach encodes the known results about the exponential generating functions for the different number of vertices.
We consider also some applications as counting hyperforests in the $k$-uniform complete hypergraph and the one complete in hyperedges of all dimensions.
Some general feature of the asymptotic regimes for large number of connected components is discussed.
\end{abstract}


\medskip


\newpage

\section{Introduction} \label{sec:intro}

In this paper we shall be concerned mainly with the problem of evaluating the weight of rooted and unrooted hyperforests in the  complete hypergraph with $n$ vertices $\overline{{\cal K}}_n$ when the weight of a hyperedge depends only on its cardinality. 
These questions are usually analyzed by using the exponential generating function and the Lagrange inversion formula~\cite{Graham_94, Wilf}, eventhough it seems that they have been posed and solved in the context of statistical mechanics~\cite{Husimi}.
But, at least in the case of ordinary graphs, the entropy of trees and rooted forests of a generic graph can be evaluated by using Kirchhoff's matrix-tree theorem. For the case of unrooted forests a solution can be obtained by the use of a novel generalization of the Kirchhoff's theorem~\cite{us_prl},
where the generating function of spanning forests in a graph, which arises as the $q \to 0$ limit of the partition function of the $q$-state Potts model \cite{Stephen_76,Wu_77,Jacobsen,alan}, can be represented as a Grassmann integral involving a quadratic (Gaussian) term together with a special nearest-neighbor four-fermion interaction.
Furthermore, this fermionic model possesses an $\OSP(1|2)$ supersymmetry.
By applying this method the classical result~\cite{Renyi,Denes} that the number of unrooted forests on the complete graph with $n$ vertices  for large $n$ behaves asymptotically as $n^{n-2} \sqrt{e}$ can be recovered~\cite{results}. But also more detailed informations. For example in~\cite{Claudia} the renormalization flow for unrooted forests on the triangular lattice has been analyzed.

A further generalization has been achieved in~\cite{noi}, where, given a hypergraph $G=(V,E)$ 
(that is, $E$ is an arbitrary collection of subsets of $V$,
 each of cardinality $\ge 2$), by exploiting the underlying $\OSP(1|2)$ supersymmetry, a class of Grassmann integrals permits an expansion in terms of spanning hyperforests. 
More precisely, let us introduce, at each vertex $i \in V$,
a pair of Grassmann variables $\psi_i$, $\psibar_i$,
which obey the usual rules for Grassmann integration \cite{Berezin,Zinn-Justin}.
For each subset $A \subseteq V$  we define the monomial  
$\tau_A=\prod_{i \in A} \psibar_i \psi_i$, and for each number $\lambda$ (in $\R$ or $\C$),
we define the Grassmann element
\begin{equation}
f_A^{(\lambda)}   \; = \;
\lambda (1-|A|) \tau_A \,+\, \sum_{i \in A} \tau_{A \smallsetminus i}
  \,-\! \sum_{\begin{scarray}
                 i,j \in A \\
                 i \neq j
              \end{scarray}}
         \!
         \psibar_i \psi_j \tau_{A \smallsetminus \{i,j\}}
  \label{eq.deff_A}
\end{equation}
and introduce a notation for the integral on all the Grassmann fields on the vertices
\begin{eqnarray}
\int \mathcal{D}_V(\psi, \psibar) & := & \prod_{i\in V} \int d\psibar_i d\psi_i \, \\
 \int \mathcal{D}_{V,\bt}(\psi, \psibar) & := &  \prod_{i\in V} \int d\psibar_i d\psi_i \, 
 \exp \left[  t_i \psibar_i \psi_i  \right] \ .
\end{eqnarray}

Given arbitrary hyperedge weights $\{ w_A \} _{A \in E}$, 
the general Grassmann integral (``partition function'')
\be
   Z =
   \int \mathcal{D}_{V,\bt}(\psi, \psibar) \,
      \exp \Biggl[ \sum_{A \in E} w_A f_A^{(\lambda)} \Biggr]
 \label{eq.Zgrass}
\ee
has a combinatorial interpretation in terms of spanning hyperforests. 

Special cases provide the generating functions for rooted and unrooted
spanning (hyper)forests and spanning (hyper)trees.
The generating function of {\em unrooted}\/ spanning hyperforests,
with a weight $w_A$ for each hyperedge $A$
and a weight $\lambda$ for each connected component is given by
\begin{eqnarray}
   \!\!
   \int \! \mathcal{D}_V(\psi, \psibar) \,
      \exp \Biggl[ \lambda \sum_{i \in V} \psibar_i \psi_i
                      \,+\,  \sum_{A \in E} w_A f_A^{(\lambda)} \Biggr]
   & = & \!\!
   \sum_{F\in \scrf(G)}
   \!
     \Biggl( \prod\limits_{A \in F} w_A \! \Biggr)
   \; \lambda^{k(F)}
         \\[2mm]
   & = &
   \lambda^{|V|}
   \!\!
   \sum_{F\in \scrf(G)}
   \!
     \Biggl( \prod\limits_{A \in F} {w_A \over \lambda^{|A|-1}} \! \Biggr)
   \;, \qquad\quad
 \label{eq.cor1.Zgrass}
\end{eqnarray}
where the sum runs over spanning hyperforests $F$ in $G$,
and $k(F)$ is the number of connected components of $F$
(note that the second equality in \reff{eq.cor1.Zgrass}
uses the following Proposition~\ref{prop.eulerineq.hypergraphs}).
If we set $w_A=1$ for all the hyperedges $A\in E$, we get as coefficient of $\lambda^p$ in the polynomial on the right hand side  of the previous equation the number  of unrooted hyperforests of the hypergraph $G$ with $p$ components.

If, on the other hand, we specialize \reff{eq.Zgrass} to $\lambda = 0$,
we obtain:
\begin{equation}
   \int \! \mathcal{D}_{V,\bt}(\psi, \psibar) \,
      \exp \Biggl[   \sum_{A \in E} w_A f_A^{(0)} \Biggr]
   \;= \!\!
   \sum_{\begin{scarray}
              F\in \scrf(G)  \\
              F = (F_1,\ldots,F_l)
         \end{scarray}}
   \!\!\!
     \Biggl( \prod\limits_{A \in F} w_A \! \Biggr)
   \prod_{\alpha=1}^l
        \Big( \sum_{i \in V(F_{\alpha})} t_i \Big)
  \;,
 \label{eq.cor2.Zgrass}
\end{equation}
where the sum runs over spanning hyperforests $F$ in $G$
with components $F_1,\ldots,F_l$,
and $V(F_\alpha)$ is the vertex set of the hypertree $F_\alpha$.
This is the generating function of {\em rooted}\/ spanning hyperforests,
with a weight $w_A$ for each hyperedge $A$
and a weight $t_i$ for each root $i$.
By taking the derivatives with respect to $t_{i_1}, \cdots, t_{i_r}$ at  $\bt =0$ we easily get the generating function of  spanning hyperforests  rooted at the vertices $i_1, \cdots, i_r$, which is
\begin{equation}
  \int \! \mathcal{D}_V(\psi, \psibar) \, (\psibar \psi)_{i_1} \cdots  (\psibar \psi)_{i_r}\,
      \exp \Biggl[ \sum_{A \in E} w_A f_A^{(0)} \Biggr] \label{rooted}
\end{equation}
where we used the shortened notation $(\psibar \psi)_{i} := \psibar_i \psi_i$.
If we now set $w_A=1$ for all the hyperedges $A\in E$, we get the number of hyperforests of the hypergraph $G$ with connected components $r$ hypertrees rooted at the vertices $i_1, \cdots, i_r$. Remark that in the case of an ordinary graph $f_A^{(0)}=f_{\{i,j\}}^{(0)}=(\psibar_i -\psibar_j)(\psi_i- \psi_j)$ is a quadratic form in the Grassmann fields, and the previous integral reduces to the evaluation of a reduced determinant of the {\em Laplacian} matrix, in agreement with the matrix-tree theorem.

In what follows we shall obtain explicit formulas for the case of the hypergraph $\overline{{\cal K}}_n$ which is complete in hyperedges of all possible cardinality, with weight $w_s$ on the hyperedges of cardinality $s$ for $s=2,\ldots,n$, that is with $w_A = w_{|A|}$. These results could in principle and in many cases have been already derived by using the standard methods of enumerative combinatorics,  that is Lagrange inversion formula in connection with the formalism of the exponential generating functions.
We hope to convince the reader that also in these cases our Grassmann formalism provides an alternative, simple and compact way to recover the total weights for rooted and unrooted hyperforests on $n$ labeled vertices, which is to say spanning on the complete hypergraph $\overline{{\cal K}}_n$.

This paper is organized as follows. In Section~\ref{sec:graph} we recall relevant notions from graph theory. 
In Section~\ref{EGF} we illustrate how, at least in the case of complete graph, the representation for the generating function of unrooted hyperforests~(\ref{eq.cor1.Zgrass}) can be deduced from that for the rooted hypertrees~(\ref{eq.cor2.Zgrass}).
In Section~\ref{sec:lemma} we collect all the explicit Grassmann integrals that will be used in the following.
In Section~\ref{sec:relation} we show the relation between our Grassmann integrals and the explicit solutions achieved by standard methods.
In Section~\ref{applications} we deal with rooted hyperforests, while Section~\ref{sec:unrooted} is devoted to unrooted hyperforests.
By restricting our general model to the case in which only one weight in nonzero, that is $w_p=\delta_{p,k}\,$, we obtain the explicit evaluation of the number of rooted and unrooted spanning hyperforests on the $k$-uniform complete hypergraphs ${\cal K}_n^{(k)}$ with $n$-vertices. These results are presented respectively in Section~\ref{sec:cgraph} and Section~\ref{sec:hyperforests}. Here we also derive a novel general simple expression for the number of unrooted hyperforests with $p$ hypertrees in terms of associated Laguerre polynomials and its asymptotic expansion for large number of vertices.
We consider also another special case, the one in which all the weights are equal, that is $w_p=1$ for all $p$, in Section~\ref{sec:complete}.  We give in Section~\ref{final} the evaluation of the number of hyperforests rooted on  $p$ vertices for the hypergraph $\overline{{\cal K}}_n$. Some conclusions are presented in Section~\ref{conclusions}.

Appendix~\ref{app:num} collects some basic features of Stirling numbers of the second kind, Bell numbers and Bell polynomials.
In Appendix~\ref{app:egf} we report, for reader convenience, the derivation of the number of (hyper-)trees in a unified way by the standard exponential generating function formalism and Lagrange inversion formula.
In Appendix~\ref{app:lag} we provide some results on the asymptotic behaviour of the associated Laguerre polynomials which are used in the main text.

\section{Graphs and hypergraphs}   \label{sec:graph}

A (simple undirected finite) {\em graph}\/ is a pair $G=(V,E)$,
where $V$ is a finite set and $E$ is a collection (possibly empty)
of 2-element subsets of $V$.
The elements of $V$ are the {\em vertices}\/ of the graph $G$,
and the elements of $E$ are the {\em edges}\/.
Usually, in a picture of a graph,
vertices are drawn as dots and edges as lines (or arcs).
Please note that, in the present definition,
loops
  (\setlength{\unitlength}{6pt}
   \begin{picture}(2.5,1)
   \put(0,0.5){\circle*{0.5}}
   \qbezier(0,0.5)(2,1.5)(2,0.5)
   \qbezier(0,0.5)(2,-0.5)(2,0.5)
   \end{picture})
and multiple edges
  (\setlength{\unitlength}{6pt}
   \begin{picture}(3.5,1)
   \put(0,0.5){\circle*{0.5}}
   \put(3,0.5){\circle*{0.5}}
   \qbezier(0,0.5)(1.5,1.5)(3,0.5)
   \qbezier(0,0.5)(1.5,-0.5)(3,0.5)
   \end{picture})
are not allowed.
We write $|V|$ (resp.\ $|E|$) for the cardinality of the
vertex (resp.\ edge) set;
more generally, we write $|S|$ for the cardinality of any finite set $S$.

A graph $G'=(V',E')$ is said to be a {\em subgraph}\/ of $G$
(written $G'\subseteq G$)
in case $V'\subseteq V$ and $E'\subseteq E$.
If $V'=V$, the subgraph is said to be {\em spanning}\/.
We can, by a slight abuse of language, identify a spanning subgraph $(V,E')$
with its edge set $E'$.

A {\em walk}\/ (of length $k \ge 0$) connecting $v_0$ with $v_k$ in $G$
is a sequence $$(v_0, e_1, v_1, e_2, v_2, \ldots, e_k, v_k)$$
such that all $v_i \in V$, all $e_i \in E$,
and $v_{i-1}, v_i \in e_i$ for $1 \le i \le k$.
A {\em cycle}\/ in $G$ is a walk in which
\begin{itemize}
   \item[(a)]  $v_0,\ldots,v_{k-1}$ are distinct vertices of $G$,
        and $v_k = v_0$
   \item[(b)]  $e_1,\ldots,e_k$ are distinct edges of $G$; and
   \item[(c)]  $k \ge 2$.\footnote{
       Actually, in a graph as we have defined it,
       all cycles have length $\ge 3$
       (because $e_1 \neq e_2$ and multiple edges are not allowed).
       We have presented the definition in this way
       with an eye to the corresponding definition for hypergraphs (see below),
       in which cycles of length 2 are possible.
}
\end{itemize}

The graph $G$ is said to be {\em connected}\/ if
every pair of vertices in $G$ can be connected by a walk.
The {\em connected components}\/ of $G$ are the
maximal connected subgraphs of $G$.
It is not hard to see that the property of being connected by a walk
is an equivalence relation on $V$,
and that the equivalence classes for this relation
are nothing other than the vertex sets of the connected components of $G$.
Furthermore, the connected components of $G$ are the induced subgraphs of $G$
on these vertex sets.\footnote{
   If $V' \subseteq V$, the {\em induced subgraph}\/ of $G$ on $V'$,
   denoted $G[V']$, is defined to be the graph $(V',E')$
   where $E'$ is the set of all the edges $e \in E$ that satisfy
   $e \subseteq V'$ (i.e., whose endpoints are in $V'$).
}
We denote by $c(G)$ the number of connected components of $G$.
Thus, $c(G)=1$ if and only if $G$ is connected.

A {\em forest}\/ is a graph that contains no cycles.
A {\em tree}\/ is a connected forest.
(Thus, the connected components of a forest are trees.)
It is easy to prove, by induction on the number of edges, that
\begin{equation}
   |E| \,-\, |V| \,+\, c(G)  \;\ge\; 0
 \label{eq.eulerineq.graphs}
\end{equation}
for all graphs, with equality if and only if $G$ is a forest.
%

In a graph $G$, a {\em spanning forest}\/ (resp.\ {\em spanning tree}\/)
is simply a spanning subgraph that is a forest (resp.\ a tree).
We denote by $\scrf(G)$ [resp.\ $\scrt(G)$]
the set of spanning forests (resp.\ spanning trees) in $G$.
As mentioned earlier, we will frequently identify
a spanning forest or tree with its edge set.

A {\em rooted tree} is a tree with a distinguished vertex called the {\em root}. A {\em rooted forest} is a graph whose connected components are rooted trees.

Hypergraphs are the generalization of graphs
in which edges are allowed to contain more than two vertices.
Unfortunately, the terminology for hypergraphs varies substantially
from author to author,
so it is important to be precise about our own usage.
For us, a {\em hypergraph}\/ is a pair $G=(V,E)$,
where $V$ is a finite set and $E$ is a collection (possibly empty)
of subsets of $V$, each of cardinality $\ge 2$.
The elements of $V$ are the {\em vertices}\/ of the hypergraph $G$,
and the elements of $E$ are the {\em hyperedges}\/
(the prefix ``hyper'' can be omitted for brevity).
Note that we forbid hyperedges of 0 or 1 vertices
(some other authors allow these).\footnote{
   Our definition of hypergraph is the same as that of
   McCammond and Meier \cite{McCammond_04}.
   It is also the same as that of Gessel and Kalikow \cite{Gessel_05},
   except that they allow multiple edges and we do not:
   for them, $E$ is a {\em multiset}\/ of subsets of $V$
   (allowing repetitions), while for us
   $E$ is a {\em set}\/ of subsets of $V$ (forbidding repetitions).
}
We shall say that $A \in E$ is a {\em $k$-hyperedge}\/
if $A$ is a $k$-element subset of $V$.
A hypergraph is called {\em $k$-uniform}\/
if all its hyperedges are $k$-hyperedges.
Thus, a 2-uniform hypergraph is nothing other than an ordinary graph.


The definitions of subgraphs, walks, cycles, connected components,
trees and forests given above for graphs were explicitly chosen
in order to immediately generalize to hypergraphs:
it suffices to copy the definitions verbatim,
inserting the prefix ``hyper'' as necessary.
The analogue of the inequality \reff{eq.eulerineq.graphs} is the following:

\begin{proposition}
   \label{prop.eulerineq.hypergraphs}
Let $G=(V,E)$ be a hypergraph.  Then
\begin{equation}
   \sum_{A \in E} (|A|-1)  \,-\, |V|  \,+\,  c(G)  \;\ge\; 0
\ef,
  \label{eq.eulerineq.hypergraphs}
\end{equation}
with equality if and only if $G$ is a hyperforest.
\end{proposition}

\noindent
Proofs can be found, for instance,
in \cite[p.~392, Proposition~4]{berge1}
or \cite[pp.~278--279, Lemma]{Gessel_05}.

Please note one important difference between graphs and hypergraphs:
every connected graph has a spanning tree,
but not every connected hypergraph has a spanning hypertree.
Indeed, it follows from Proposition~\ref{prop.eulerineq.hypergraphs}
that if $G$ is a $k$-uniform hypergraph with $n$ vertices,
then $G$ can have a spanning hypertree only if $k-1$ divides $n-1$.
Of course, this is merely a necessary condition, not a sufficient one!

The hypergraph ${\cal K}_n^{(k)}$ has $|V|=n$ vertices and is complete in the $k$-hyperedges, in the sense that it is $k$-uniform and for all choices of $k$ different vertices $i_1,\dots, i_k$ in $V$ the hyperedge $\{i_1, \ldots, i_k\}$ belongs to the set of its hyperedges.

The hypergraph\footnote{We don't use the symbol ${\cal K}_n$ because this is usually used for the complete graph ${\cal K}_n^{(2)}$ with $n$ vertices.}  $\overline{{\cal K}}_n$ has $|V|=n$ vertices and is complete in the $k$-hyperedges for all $2\leq k \leq n$, so that $E\left(\overline{{\cal K}}_n \right) = \bigcup_{k=2}^n \, E\left( {\cal K}_n^{(k)} \right)$.

\section{Exponential generating function for hypertrees and hyperforests} \label{EGF}

Let us consider the complete hypergraph $\overline{{\cal K}}_n$ for every $n$, with general hyperedge-weights $w_A$ which vary only with the cardinality of the hyperedge $A$, i.e. $w_A = w_{|A|}$. The $k$-uniform complete hypergraph ${\cal K}_n^{(k)}$ corresponds to the case in which the only non-vanishing weight is $w_k$.

Let $t_n$ be the total weight of rooted hypertrees in the case of $n$ vertices $|V|=n$, $\bw = \{w_k\}_{k\ge 2}$ and let 
\be
T(z) = T(z,\bw)  := \sum_{n\ge 0} t_n(\bw)\, \frac{z^n}{n!}
\ee
the exponential generating function for the sequence $\{t_n\}$.

The exponential generating function for rooted hyperforests is therefore $e^{\,t\,T(z)}$, where $t$ counts the number of connected components.

In the case of complete hypergraphs  we can also consider the exponential generating function for unrooted trees
\be
U(z) = U(z,\bw)   := \sum_{n\ge 0} u_n(\bw) \, \frac{z^n}{n!}
\ee
where $u_n$ is the weight of unrooted trees in the case of $n$ vertices $|V|=n$.
Of course as the root of a trees on $n$ vertices can be chosen in $n$ ways 
\be t_n = n\, u_n \ee
and therefore
\be
T(z) = z\, \frac{d}{dz} U(z)
\ee
and conversely
\be
U(z)  = \int_0^z\, \frac{d \omega}{\omega}\,T(\omega)\, . \label{intw}
\ee
By using  a recursion relation, as it is done in Appendix~\ref{app:egf}, we see that the exponential generating function for rooted hypertrees satisfies the relation
\be
T(z) = z\, \exp\left[ \sum_{k\ge 2} w_k \,\frac{T(z)^{k-1}}{(k-1)!} \right]\label{i}
\ee
therefore
\be
z = T\,\exp\left[ - \sum_{k\ge 2} w_k \,\frac{T^{k-1}}{(k-1)!} \right]
\ee
and by changing variables from $\omega$ to $T(\omega)$ in the integral in~(\ref{intw}) we easily get
\be
U(z) = T(z) + \sum_{k\ge 2} w_k\, (1-k)\,\frac{T(z)^{k}}{k!}\label{T}
\ee
that is the exponential generating function for unrooted hypertrees can be expressed in terms of the exponential generating function of rooted hypertrees~\cite{Husimi}. Furthermore, the exponential generating function for unrooted hyperforests, with parameter $\lambda$ to count the number of connected components, is simply $e^{\,\lambda\,U(z)}$ and this can also be expressed in terms of the exponential generating function of rooted hypertrees by means of~(\ref{T}).

Let us use these relations  in order to re-obtain, at least in the case considered here of the complete hypergraph, the generating function of unrooted hyperforests in the Grassmann representation from the
generating function of rooted hyperforests.

Formula~(\ref{eq.cor2.Zgrass}) for the generating function of unrooted hyperforests for $\overline{{\cal K}}_n$, at $t_i=t$ for every vertex, means that 
\be
n!\,\left[z^n\right]\, e^{t\,T(z)} =  \int \mathcal{D}_n(\psi, \psibar) \,\exp\left[t\, (\psibar, \psi)  + \sum_{A\in E} w_{|A|} f_A^{(0)}\right] \label{S0}
\ee
where we shortly denoted 
\be
\mathcal{D}_n(\psi, \psibar) := \mathcal{D}_{V\left(\overline{{\cal K}}_n\right)}(\psi, \psibar) = \prod_{i=1}^n \int d\psibar_i d\psi_i \label{defD}
\ee
and
\be
(\psibar, \psi) := \sum_{i \in V\left(\overline{{\cal K}}_n\right)} \left(\psibar \psi\right)_i = 
\sum_{i=1}^n \psibar_i \psi_i \ef
\ee
And, if $f(z)$ can be expanded in powers of $z$,  $[z^n]\,f(z)$ is the coefficient of $z$ in the expansion.

It follows that ,for every power $r$, the coefficient of $t^r$ is equal to
\be
n!\,\left[z^n\right]\, T(z)^r =  \int \mathcal{D}_n(\psi, \psibar) \,(\psibar, \psi)^r \,\exp\left[ \sum_{A\in E} w_ {|A|} f_A^{(0)}\right]
\ee
and therefore for each function $L$ defined by a formal power series
\be
n!\,\left[z^n\right]\, L[T(z)] =  \int \mathcal{D}_n(\psi, \psibar) \,L[(\psibar, \psi)] \,\exp\left[ \sum_{A\in E} w_{|A|} f_A^{(0)}\right]\, .\label{ris}
\ee

Now, the exponential generating function for unrooted hyperforests is $e^{\lambda\, U(z)}$, where $\lambda$ counts the hypertrees in the hyperforests and we know by (\ref{eq.cor1.Zgrass}) that
\be
n!\,\left[z^n\right]\, e^{\lambda\, U(z)} = \int \mathcal{D}_n(\psi, \psibar)\,\exp\left\{ \lambda \,(\psibar, \psi) + \sum_{A\in E} w_{|A|} f_A^{(\lambda)}\right\}
\ee
but
\be
\sum_{A\in E} w_{|A|} f_A^{(\lambda)} = \sum_{A\in E} w_{|A|} \left[ \lambda\, \left(1 - |A|\right)\,\tau_A + f_A^{(0)}\right]
\ee
and
\be
\sum_{A\in E} w_{|A|} \, \left(1 - |A|\right)\,\tau_A  = \sum_{k\ge 2}\, w_k\,  \left(1 - k\right)\, \sum_{A: |A|=k}  \tau_A 
= \sum_{k\ge 2}\, w_k \,(1-k)  \,\frac{(\psibar, \psi)^{k}}{k!}
\ee
so that
\begin{multline}
n!\,\left[z^n\right]\, e^{\lambda\, U(z)} =  \\
\int \mathcal{D}_n(\psi, \psibar)\,\exp\left\{ \lambda\,\left[\,(\psibar, \psi) + \sum_{k\ge 2} w_k\, (1-k)\,\frac{(\psibar, \psi)^{k}}{k!} \right] + \sum_{A\in E} w_{|A|} f_A^{(0)}\right\} \label{T0}
\end{multline}
But this is exactly formula~(\ref{ris}) when
\be
L(y) = e^{\,\lambda\, K(y)} \label{f:L}
\ee
with
\be
K(y) := y  + \sum_{k\ge 2} w_k\, (1-k)\,\frac{y^{k}}{k!} \label{f:K}
\ee
which is such that $U(z) = K[T(z)]$ by~(\ref{T}).

\section{Useful Lemmas on Grassmann integrals}   \label{sec:lemma}

In the following we shall make use of very simple results for Grassmann integrals.

\begin{lemma}
 \label{int}
Let $|V|=n$ be the number of vertices, then
$$\int \mathcal{D}_n(\psi, \psibar) \,\frac{(\psibar,  \psi)^s}{s!} = \delta_{s,n} $$ \ .
\end{lemma}

\proof
It trivially follows from induction in $n$. \qed

We soon derive, by expansion in powers, that
\begin{corollary}
 \label{coroll}
Let $g$ be  a generic function of the scalar product, that is a polynomial as the scalar product is nilpotent of degree $n$, then
$$
\int \mathcal{D}_n(\psi, \psibar) \, g\left( (\psibar, \psi) \right) = n!\,\left[z^n\right] \,g(z) = \frac{n!}{2\,\pi\,i}\,\oint \frac{dz}{z^{n+1}}\,g(z)\label{contour}
$$
where  the contour integral is performed in the complex plain constrained to encircle the origin.
\end{corollary}

These are the ingredients to observe that
\begin{lemma}
 \label{roots}
 Let $|V|=n$ be the number of vertices, $g$ a generic function, then 
\begin{align*}
\int \mathcal{D}_n(\psi, \psibar) \, (\psibar \psi)_{i_1} \cdots  (\psibar \psi)_{i_r}\, g\left( (\psibar, \psi) \right)  & =  \frac{(n-r)!}{n!} \int \mathcal{D}_n(\psi, \psibar) \, (\psibar, \psi)^r\, g\left( (\psibar, \psi) \right) \\
& = (n-r)!\,\left[z^{n-r}\right] \,g(z)
\end{align*}
\end{lemma}

\proof
By integrating over $\psibar_{i_1}, \psi_{i_1}, \cdots, \psibar_{i_r}, \psi_{i_r}$ on the left hand side we get an integral of the form used in the previous Lemma, where both the integration measure and the scalar product were restricted on the remaining $n-r$ vertices, so that
$$\int \mathcal{D}_{n-r}(\psi, \psibar) \, g\left( (\psibar, \psi) \right) =  (n-r)!\,\left[z^{n-r}\right] \,g(z)\ef.
$$
By expanding instead on the right hand side we get 
$$\sum_{s\ge 0} \frac{(n-r)!}{n!} \int \mathcal{D}_n(\psi, \psibar) \, (\psibar, \psi)^{r+s} \, [z^s] \,g(z) = \ (n-r)!\,\left[z^{n-r}\right] \,g(z)$$
and we get our result by using the previous Lemma~\ref{int}. \qed

Let $J$ the matrix with unit entries for each $i,j\in V$
\be
J_{ij} = 1 
\ef.
\ee
Our common tool is the following

\begin{lemma}
  \label{lemma.integral}
   Let $|V|=n$ be the number of vertices, $g$ and $h$ generic function, then 
  \begin{eqnarray*}
 \lefteqn{ \int \mathcal{D}_n(\psi, \psibar)\, (\psibar, \psi)^r\,
 e^{h\left( (\psibar, \psi) \right) + (\psibar, J \psi) g\left( (\psibar, \psi) \right)} =}\\
&=&  \ \int \mathcal{D}_n(\psi, \psibar) \, (\psibar, \psi)^r\,
 e^{h\left( (\psibar, \psi) \right)} \left[ 1 +  (\psibar, \psi)\, g\left( (\psibar, \psi) \right) \right]
 \end{eqnarray*}
\end{lemma}
 
 \proof
 Let us expand the second part of the exponential
   \begin{eqnarray*}
 \lefteqn{ \int \mathcal{D}_n(\psi, \psibar) \, (\psibar, \psi)^r\,
 e^{h\left( (\psibar, \psi) \right)} \sum_{s} \frac{ (\psibar, J \psi)^s}{s!} \, g\left( (\psibar, \psi) \right)^s=}\\
&=&   \int \mathcal{D}_n(\psi, \psibar) \, (\psibar, \psi)^r\,
 e^{h\left( (\psibar, \psi) \right)} \left[ 1 +  (\psibar, J \psi) \, g\left( (\psibar, \psi) \right) \right] \\
&=&   \int \mathcal{D}_n(\psi, \psibar) \, (\psibar, \psi)^r\,
 e^{h\left( (\psibar, \psi) \right)} \left[ 1 +  (\psibar, \psi)\, g\left( (\psibar, \psi) \right) \right]
  \end{eqnarray*}
 because all higher powers of $(\psibar, J \psi)$ vanish. We get the final line because in the rest of the integral for each $i$ the field $\psibar_i$ is always multiplied by the companion $\psi_i$ and thus the only contribution in $(\psibar, J \psi)$ comes from the diagonal part, that is $(\psibar, \psi)$.~ \qed

\section{Relation with previous approaches}   \label{sec:relation}

In virtue of our Lemmas, the Grassmann integrals for the generating functions of rooted and unrooted hyperforests at fixed number of vertices can be expressed as a unique contour integral of a complex variable. In this section we will show the change of variables which explicitly maps those integrals into the coefficient of the corresponding exponential generating function in the number of vertices, without using the Lagrange inversion formula.

The sum on all the edges appears in both main formulas~(\ref{S0}) and~(\ref{T0})  and in our model it becomes
\begin{eqnarray}
\sum_{A\in E} w_{|A|} f_A^{(0)}  &=& \sum_{k\ge 2} w_{k} \sum_{A: |A|=k} f_A^{(0)}\nonumber \\
 &=& \sum_{k\ge 2} w_{k} \,\left[(n-k+1)  \frac{(\psibar,\psi)^{k-1}}{(k-1)!} - (\psibar,(J-I) \psi)  \frac{(\psibar,\psi)^{k-2}}{(k-2)!}\right]\nonumber  \\ 
 &=&  \sum_{k\ge 2} w_{k}  \,\left[ n  \frac{(\psibar,\psi)^{k-1}}{(k-1)!} - (\psibar,J \psi)  \frac{(\psibar,\psi)^{k-2}}{(k-2)!}\right]  \label{eq.sumf_A}
\end{eqnarray}
and according to Lemma~\ref{lemma.integral}, for any function $h$ of the scalar product $(\psi, \psibar)$
\begin{multline}
\int \mathcal{D}_n(\psi, \psibar) 
 \,h\left( (\psibar, \psi) \right)\,\exp\left[ \sum_{A\in E} w_{|A|} f_A^{(0)} \right] =\\
 \int \mathcal{D}_n(\psi, \psibar) 
 \,h\left( (\psibar, \psi) \right)\,\exp\left[ n \sum_{k\ge 2} w_{k}  \,  \frac{(\psibar,\psi)^{k-1}}{(k-1)!} \right]\, \left[ 1 - \sum_{k\ge 2} w_{k}  \, \frac{(\psibar,\psi)^{k-1}}{(k-2)!}\right] 
\end{multline}
so the Grassmann integrals reduces to what has been formally obtained in Corollary~\ref{coroll} and we have for~(\ref{S0})
\begin{multline}
n!\,\left[z^n\right]\, e^{\,t\,T(z)}
 =   \int \mathcal{D}_n(\psi, \psibar)  \, \left[ 1 - \sum_{k\ge 2} w_{k}  \, \frac{(\psibar,\psi)^{k-1}}{(k-2)!}\right]\\
\,\times\,\exp\left[t\, (\psibar, \psi) +  n \sum_{k\ge 2} w_{k}  \,  \frac{(\psibar,\psi)^{k-1}}{(k-1)!} \right]\\
=  \frac{n!}{2\,\pi\, i} \oint \frac{d\xi}{\xi^{n+1}}   \, \left[ 1 - \sum_{k\ge 2} w_{k}  \, \frac{\xi^{k-1}}{(k-2)!}\right]
\,\exp\left[t\, \xi +  n \sum_{k\ge 2} w_{k}  \,  \frac{\xi^{k-1}}{(k-1)!} \right]
\end{multline}
which is nothing but
\be
\left[z^n\right]\, e^{\,t\,T(z)} = \frac{1}{2\,\pi\, i} \oint \frac{dz}{z^{n+1}}  \,e^{\,t\,T(z)}
\ee
with the change of variables~(\ref{i}) with $T(z)=\xi$, as
\be
\frac{dz}{z} = \frac{d\xi}{\xi}\,\left[ 1 - \sum_{k\ge 2} w_{k}  \, \frac{\xi^{k-1}}{(k-2)!}\right]\ef.
\ee
Analogously for~(\ref{T0})
\begin{align}
n!\,\left[z^n\right]\, e^{\,\lambda\,U(z)}
&  =  \int \mathcal{D}_n(\psi, \psibar)  \, \left[ 1 - \sum_{k\ge 2} w_{k}  \, \frac{(\psibar,\psi)^{k-1}}{(k-2)!}\right]\nonumber\\
& \quad\,\times\,\exp\left[\lambda\, \left( (\psibar, \psi) \,+ \,\sum_{k\ge 2}w_{k}  \, (1-k)\,\frac{(\psibar, \psi)^k}{k!}\right) +  n \sum_{k\ge 2} w_{k}  \,  \frac{(\psibar,\psi)^{k-1}}{(k-1)!} \right]\nonumber\\
& =  \frac{n!}{2\,\pi\, i} \oint \frac{d\xi}{\xi^{n+1}}   \, \left[ 1 - \sum_{k\ge 2} w_{k}  \, \frac{\xi^{k-1}}{(k-2)!}\right]\nonumber\\
& \quad\,\times\,\exp\left[\lambda\, \left(\xi\,+ \,\sum_{k\ge 2}w_{k}  \, (1-k)\,\frac{\xi^k}{k!}\right)  +  n \sum_{k\ge 2} w_{k}  \,  \frac{\xi^{k-1}}{(k-1)!} \right]
\end{align}
which, by using the same change of variables, is nothing but
\begin{align}
\left[z^n\right]\, e^{\,\lambda\,U(z)} & = \frac{1}{2\,\pi\, i} \oint \frac{dz}{z^{n+1}}  \,e^{\,\lambda\,U(z)}\\
& =
\frac{1}{2\,\pi\, i} \oint \frac{dz}{z^{n+1}}  \,\exp\left\{\,\lambda\, \left[T(z)\,+ \,\sum_{k\ge 2}w_{k}  \, (1-k)\,\frac{T(z)^k}{k!}\right]\right\}
\end{align}

\section{Rooted hyperforests} \label{applications}
 
Let us begin with the evaluation of~(\ref{rooted}), that is the case $\lambda=0$ which evaluates the weight of rooted hyperforests on $r$ vertices $i_1,\cdots, i_r$, which on the  complete hypergraph $\overline{{\cal K}}_n$ does not depend on the particular choice of the vertices, and we denote this weight by $v_{n,r}$.   We have 
\begin{eqnarray*}
v_{n,r} = v_{n,r}(\bw)  &=& \int \mathcal{D}_n(\psi, \psibar) \, (\psibar \psi)_{i_1} \cdots  (\psibar \psi)_{i_r}\,
\\
&  &\qquad \exp {\left[n \sum_{k\ge 2} w_k \frac{ (\psibar, \psi)^{k-1} }{(k-1)!} -  \sum_{k\ge 2} w_k (\psibar, J \psi)  \frac{ (\psibar, \psi)^{k-2} }{(k-2)!}\right]}\\ 
&=&  \frac{(n-r)!}{n!}   \int \mathcal{D}_n(\psi, \psibar) \,(\psibar,  \psi)^r \,
 \left[ 1 -   \sum_{k\ge 2} w_k  \frac{ (\psibar, \psi)^{k-1} }{(k-2)!} \right] \\
 & & \qquad  \exp \left[{ n \sum_{k\ge 2} w_k \frac{ (\psibar, \psi)^{k-1} }{(k-1)!} } \right]
 \end{eqnarray*}
Of course there are ${n \choose r}$ different choiches for $r$ different vertices, therefore, if we denote by 
\be
E_n(t; \bw) = n!\, [z^n]\, e^{t\,T(z)}
\ee
the generating function of rooted hyperforests on $n$ vertices, its Grassmann representation is
\be
E_n(t; \bw) = \int \mathcal{D}_n(\psi, \psibar)  \,
 \left[ 1 -   \sum_{k\ge 2} w_k  \frac{ (\psibar, \psi)^{k-1} }{(k-2)!} \right]  e^{ t\,(\psibar, \psi) + n\, \sum_{k\ge 2} w_k \frac{ (\psibar, \psi)^{k-1} }{(k-1)!} }\, . \label{ent}
\ee
The expansion in power series of $t$
\be
E_n(t; \bw) = \sum_{r\ge 0} t_{n,r}(\bw)\,t^r
\ee
provides the total weight of rooted hyperforests  with $r$ connected components 
\be
t_{n,r} =  t_{n,r}(\bw) = n!\, [z^n]\,[t^r]\, e^{t\,T(z)} = [t^r]\, E_n(t; \bw)
\ee
then
\be
t_{n,r} =   {n \choose r} \,v_{n,r} =    \int \mathcal{D}_n(\psi, \psibar) \,\frac{(\psibar,  \psi)^r}{r!} \,
 \left[ 1 -   \sum_{k\ge 2} w_k  \frac{ (\psibar, \psi)^{k-1} }{(k-2)!} \right]  e^{ n \sum_{k\ge 2} w_k \frac{ (\psibar, \psi)^{k-1} }{(k-1)!} }
\ee
while then the total weight of rooted hyperforests
\be
E_n(\bw) := E_n(1; \bw) = \sum_{r\ge 0} t_{n,r}(\bw)
\ee
is given by the generating function at $t=1$.

Let us now introduce the function
\be
\theta (x, y; \bw) := \exp\left[ x \sum_{k\ge 2} w_k \,  \frac{y^{k-1}}{(k-1)!} \right] := \sum_{s\ge 0} P_s(x; \bw) \,\frac{y^s}{s!} \label{teta}
\ee
which is the exponential generating function for the exponentials $P_s(x; \bw)$ in the variable $x$, which varies with the choice of the weights $\bw$. We  recognize that
\be
 \sum_{k\ge 2} w_k  \frac{ y^{k-1} }{(k-2)!}\, \theta (x, y; \bw) =
 \frac{y}{x} \frac{\partial}{\partial y} \theta (x, y; \bw) = \frac{1}{x}\,\sum_{s\ge 1} P_s(x; \bw) \,\frac{y^s}{(s-1)!}
\ee
Therefore the integral in~\reff{ent} can be re-expressed by using
\be
e^{ t \, y }\, \theta (x, y; \bw) =  \sum_{s\ge 0}\,\sum_{r\ge 0} P_s(x; \bw) \,t^r\,\frac{y^{r+s}}{r!\, s!} 
\ee
and
\be
 \sum_{k\ge 2} w_k  \frac{ y^{k-1} }{(k-2)!}\, e^{ t \, y }\, \theta (x, y; \bw)    = \sum_{s\ge 0}\,\sum_{r\ge 0} P_s(x; \bw) \,t^r\,\frac{y^{r+s}}{r!\, s!} \, \frac{s}{x} \label{primaf}
\ee
The same expression could be written also with the help of the derivative with respect to the variable $t$, let  $D = \frac{\partial}{\partial t}$, then
\begin{align}
 \sum_{k\ge 2} w_k  \frac{ y^{k-1} }{(k-2)!}\,&e^{ t \, y }\, \theta (x, y; \bw)  =  \\
 & =   \sum_{k\ge 2}\,  \frac{  w_k }{(k-2)!}\,D^{k-1}\,e^{ t \, y }\, \theta (x, y; \bw)  \\
& =   \sum_{k\ge 2}\,  \frac{  w_k }{(k-2)!}\, D^{k-1}\,\sum_{s\ge 0}\,\sum_{r\ge 0} P_s(x; \bw) \,t^r\,\frac{y^{r+s}}{r!\, s!} \\
 & = \sum_{s\ge 0}\, P_s(x; \bw) \sum_{r\ge k-1}\,\frac{y^{r+s}}{ s!} \, \sum_{k\ge 2}\,  \frac{  w_k }{(k-2)!}\,
 \frac{1}{\left[r-(k-1) \right]!} \, t^{r-(k-1)}\\
& = \sum_{k\ge 2}\, \sum_{s\ge k-1}\, P_{s-(k-1)}(x; \bw) \sum_{r\ge 0}\, t^{r} \,\frac{y^{r+s}}{r!\,\left[s-(k-1) \right]!} \,  \frac{  w_k }{(k-2)!} \label{secondaf}
 \end{align}
 so that by comparing term by term in \reff{primaf} and \reff{secondaf} we recover a recursion relation for the polynomials $P_s(x, \bw)$ 
 \be
 P_s(x; \bw) \,=\,  x\, \sum_{k\ge 2}\,w_k\, {s-1 \choose k-2}\, P_{s-(k-1)}(x; \bw) \, .\label{id:P}
\ee

In terms of the polynomials   $P_s(x, \bw)$ we soon get for the generating function of rooted hyperforests 
\begin{eqnarray}
E_n(t; \bw)  &=&  \sum_{s\ge 0}\,\sum_{r\ge 0} P_s(n; \bw)\, t^r\,  \int \mathcal{D}_n(\psi, \psibar) \, \frac{(\psibar,  \psi)^{r+s}}{r!\,s!} \, \left[ 1 -    \frac{s }{n} \right] \\
&=& \sum_{r\ge 1}\,  {n-1 \choose r-1} \,P_{n-r}(n; \bw) \,t^r 
\end{eqnarray}

Therefore the total weight of rooted hyperforests is
\be
E_n(\bw) = \sum_{r\ge 1}\,  {n-1 \choose r-1} \,P_{n-r}(n; \bw)
\ee
and the total weight of rooted hyperforests with $r$ hypertrees is
\be
t_{n,r}
 =   {n-1 \choose r-1} \,P_{n-r}(n; \bw)   
 \label{eq:t}
 \ee
 from which in particular we obtain for $r=0$ 
 \be
 t_{n,0} = 0 
 \ee
for all choices of the weights $\bw$, a generalization of what occurs for the case of ordinary trees because the determinant of the weighted Laplacian on the graph is always vanishing.

Also, as $P_0(x; \bw) = 1$ for all choices of the weights $\bw$, of course
\be
t_{n,n}=1
\ee
as there is only one possible hyperforest with $n$ hypertrees, the trivial one in which each hypertree is a vertex.

The weight of rooted hypertrees $t_n$ is given by the case $r=1$
\be
t_n := t_{n,1} =  \,   P_{n-1} (n; \bw) \, .
\ee

A more explicit expression for the polynomials $P_s(x; \bw)$ is obtained by expanding the exponential in the definition~\reff{teta}
 \begin{eqnarray*}
P_s(x; \bw) & = & s!\,  [y^{s}] \, \theta (x, y; \bw) \\
& = & 
 s!\,\prod_{j\ge 2} \sum_{l_j}  \frac{1}{l_j!} \left( \frac{x  w_j}{(j-1)!}\right)^{l_j} \,y^{l_j(j-1)} \\
&=& 
s!\, \sum_{\{l_j\}} \delta_{s,\sum_{j\ge 2} l_j (j-1)} \left[ \prod_{j\ge 2}  \frac{1}{l_j!} \left( \frac{x  w_j}{(j-1)!}\right)^{l_j}\right] 
\end{eqnarray*}
so that if we define the coefficients $p_{s,l}(\bw)$ by
\be
P_s(x; \bw) = \sum_{l\ge 0} p_{s,l}(\bw)\,x^l
\ee
we get
\begin{eqnarray*}
 p_{s,l} \, =  \,  p_{s,l}(\bw) & = & s!\,  [y^{s}] \, [x^l]\,\theta (x, y; \bw) \\
& = &  s!\,  \sum_{\{l_j\}} \delta_{l,\sum_{j\ge 2} l_j } \delta_{s,\sum_{j\ge 2} l_j (j-1)} \left[ \prod_{j\ge 2}  \frac{1}{l_j!} \left( \frac{ w_j}{(j-1)!}\right)^{l_j}\right] \, .
\end{eqnarray*}
In order to understand the constraint which is imposed in the sum on the coefficients $l_j$'s,  remember that from Proposition~\ref{prop.eulerineq.hypergraphs}, if $l_j$ is the number of hyperedges of cardinality $j$, $n$ is the number of vertices and $r$ is the number of connected components, which in our case is the number of hypertrees
\be
0 = \sum_{A\in E} (|A|-1) - |V| + c(G) =  \sum_{j\ge 2} l_j (j-1) - n +r \label{Euler}
\ee
and this is exactly the constraint which is imposed. The number $l$ is instead $n_E$  the total number of hyperedges.

\subsection{On the $k$-uniform complete hypergraph}   \label{sec:cgraph}

In the $k$-uniform complete hypergraph ${\cal K}_n^{(k)}$ the hyperedges are all the subsets $A\subset V$ of  $k$ vertices: $|A|=k$. This is therefore the particular case of our model in which if we introduce the
vectors $\ve_k$ such that their components are
\be
(\ve_k)_s   \,=\, \delta_{ks}
\ee
we have weights
\be
\bw = w\, \ve_k
\ee 
and as we wish to count configurations we have to set $w=1$ so that in the general formulas $w_k=1$ and all the others weights for the hyperedges have to be set to zero. We have 
\be
\theta (x, y; \ve_k) = \exp\left[\,x  \frac{y^{k-1}}{(k-1)!} \right] 
\ee
and therefore
\be
P_s(x; \ve_k) = \begin{cases}
\frac{s!}{ \left(\frac{s}{k-1}\right)! [(k-1)!]^\frac{s}{k-1} } \,  x^\frac{s}{k-1} & \hbox{if } s = l (k-1)\, \hbox{for integer } l \\
0 & \hbox{otherwise}
\end{cases}
\ee
which satisfy the recursion relation~\reff{id:P} which for $\bw = \ve_k$  takes the form
\be
P_s(x; \ve_k) =  x \, {s-1\choose k-2}\, P_{s-(k-1)}(x; \ve_k)\, .
\ee

We easily get that
\be
p_{s,l} (\ve_k)= \begin{cases}
\frac{s!}{ l! [(k-1)!]^l}  & \hbox{if } s = l (k-1)\, \hbox{for integer } l \\
0 & \hbox{otherwise}
\end{cases}
\ee
On ${\cal K}_n^{(k)}$, the numbers $n_E=l$ of  hyperedges and the number of connected components $c(G)=r$ are related by~\reff{Euler}
\be
l \, (k-1) - n + r =0
\ee
that is
\be
n_E = l = \frac{n-r}{k-1}
\ee
is the number of hyperedges (of degree $k$).

For the number of rooted hyperforests with $r$ hypertrees  on the $k$-uniform complete hypergraph ${\cal K}_n^{(k)}$,  we have when $n-r$ can be divided by $k-1$
\begin{eqnarray}
t_{n,r}(\ve_k) & = & {n-1 \choose r-1}\,P_{n-r}(n; \ve_k) \\
& = &
{n-1 \choose r-1}\ \frac{(n-r)!}{ \left(\frac{n-r}{k-1}\right)! [(k-1)!]^\frac{n-r}{k-1} }
  \,  n^\frac{n-r}{k-1} \label{pris} \\
& = & \,{(k-1) n_E +r-1 \choose r-1}\, \frac{\left[(k-1) n_E\right]!}{n_E! \left[(k-1)!\right]^{n_E}} \left[ (k-1) n_E +r \right]^{n_E}\label{ne}
 \end{eqnarray}
where the prefactor in \reff{pris}
\be
 \frac{(n-r)!}{ \left(\frac{n-r}{k-1}\right)! [(k-1)!]^\frac{n-r}{k-1} }
\ee
is exactly the number of ways in which $n-r$ vertices can be divided into $(n-r)/(k-1)$ groups of  $k-1$ elements and in \reff{ne} we have replaced the dependence from the number of vertices $n$ with that from the number of hyperedges $n_E$.
 
In the case of  simple graphs ($k=2$)  it follows that
\be
t_{n,r}(\ve_2)   = {n-1 \choose r-1}\,n^{n-r}
\ee
which at $r=1$ provides the well-known result by Cayley about the number $u_n^{(2)}$ of spanning unrooted  trees on the complete graph with $n$ vertices
\be
u_n(\ve_2) = \frac{t_n(\ve_2)}{n} = \, n^{n-2}\, .
\ee
Also
\be
E_n(t; \ve_2) = \sum_{r\ge 1}  {n-1 \choose r-1}\,n^{n-r}\, t^r = t\,(n+t)^{n-1}
\ee
which could be obtained by direct evaluation as
\be
E_n(t; \ve_2) =  \, \int \mathcal{D}_n(\psi, \psibar)  \, \left[ 1 -    (\psibar, \psi)  \right]  e^{ (t+n) \,(\psibar, \psi) }
=  \, (t+n)^n \left[ 1 - \frac{n}{n+t} \right] \label{e2}
\ee
This relation says at $t=1$  that the total number of rooted forests is
\be
E_n (\ve_2) = (n+1)^{n-1}\, .
\ee
In this simple case also the whole generating function can be expressed in terms of the generalized exponential~\cite{Graham_94}  (the usual exponential is at $\alpha=0$)
\be
{\cal E}_\alpha (z) := \sum_{n\ge 1}\,\left(\alpha n + 1\right)^{n-1}\, \frac{z^n}{n!} \label{expt}
\ee
which satisfies
\be
{\cal E}_\alpha (z)^{-\alpha}\,\ln {\cal E}_\alpha (z) = z \qquad {\cal E}_\alpha (z) = {\cal E} (\alpha z)^\frac{1}{\alpha} \label{propexpt}
\ee
where $ {\cal E} (z)$ is a shorthand for ${\cal E}_1 (z)$.
Indeed
\begin{align}
e^{t\,T(z)} = &\, \sum_{n\ge 1}\, E_n(t; \ve_2)\, \frac{z^n}{n!} = \sum_{n\ge 1}\,\left(\frac{n}{t} + 1\right)^{n-1}\, \frac{(t\,z)^n}{n!}  \nonumber \\
= & \,{\cal E}_\frac{1}{t} (t\,z) = {\cal E} (z)^t = e^{t\,z\,{\cal E}(z)}
\end{align}


\subsection{On the complete hypergraph}   \label{sec:complete}

We shall consider here the complete hypergraph $\overline{{\cal K}}_n$ when all the hyperedge-weights $w_d$ are set to one, that is
\be
\bw = \bone
\ee
where $\bone$ is the vector with $1$ on all components.
We have 
\be
\theta ( x, y; \bone) := \exp\left[ \,x\, \left(e^y -1\right)\right] = \sum_{s\ge 0} b_s(x)  \frac{y^s}{s!} 
\ee
where $b_s(x)$ are the Bell polynomials,  see Appendix~\ref{app:num}, and therefore
\be
P_s(x; \bone) = b_s(x) = \sum_{l\ge 0}  \left\{ {s \atop l} \right\}\,x^{l}
\ee
so that
\be
p_{s,l}(\bone) = \left\{ {s \atop l} \right\}
\ee
where $\left\{ {s \atop l} \right\}$ is a Stirling number of the second kind, and it is the number of ways to partition a set of cardinality $s$ into $l$ nonempty subsets. 

The recursion relation~\reff{id:P} becomes here
\be
b_s(x) = x\,\sum_{k\ge 1} {s-1 \choose k-1} \,b_{s-k}(x) \label{id:b}
\ee

The number of rooted hyperforests with  $r$ hypertrees  on the $k$-uniform complete hypergraph $\overline{{\cal K}}_n$ is therefore
\be
t_{n,r}(\bone) =  {n -1\choose r-1}\,   b_{n-r}(n) = {n -1\choose r-1}\,   \sum_{n_E\ge 0} n^{n_E} \left\{ {n-r \atop n_E} \right\} \ef. \label{tbone}
\ee
and the total number of rooted hyperforests is
\be
E_n(\bone) = \sum_{r\ge 1} t_{n,r}(\bone) = \sum_{k\ge 1}  {n-1 \choose k-1} \, b_{n-k}(n)  = \frac{b_n(n)}{n}
\ee
because of \reff{id:b} for $x=n$.

\section{Unrooted hyperforests}\label{sec:unrooted}

According to our general formula the generating function for unrooted hyperforests on $n$ vertices is given by the Grassmann integral
\begin{multline}
F_n(\lambda; \bw) := n! \,[z^n]\,e^{\lambda U(z)} = \\
 =  \int \mathcal{D}_n(\psi, \psibar) \, 
 \exp\left\{\lambda \left[(\psibar, \psi) + \sum_{k\ge 2} w_k \,(1-k)\frac{ (\psibar, \psi)^{k} }{k!}\right]\right\}  \,  \\ \times \,\exp \left[n \sum_{k\ge 2} w_k \,\frac{ (\psibar, \psi)^{k-1} }{(k-1)!} - (\psibar, J \psi) \,\sum_{k\ge 2} w_k \, \frac{ (\psibar, \psi)^{k-2} }{(k-2)!}\right] \label{Fn}
\end{multline}
which we expand in $\lambda$
\be 
F_n(\lambda; \bw) =  \sum_{p=0}^\infty\, u_{n,p}(\bw) \, \lambda^p
\ee
where $u_{n,p}(\bw)$ is the total weight of unrooted hyperforests with $p$ hypertrees.

We find convenient to introduce the polynomials $\Pi_s(\lambda; \bw)$ and the coefficients $\pi_{s,r}(\bw)$ according to
\be
\exp \left[ \lambda \left(  y  + \sum_{k\ge 2} w_k\, (1-k)\,\frac{y^{k}}{k!} \right) \right] \, = \,
\sum_{s\ge 0} \, \Pi_s(\lambda; \bw) \,\frac{y^s}{s!} \, = \, \sum_{s\ge 0} \,  \sum_{p\ge 0}\, \pi_{s,p}(\bw) \,\lambda^p\, \frac{y^s}{s!} \label{def:Pi}
\ee
It soon follows that
\begin{align}
F_n(\lambda; \bw)  =    \sum_{s\ge 1}\, & \Pi_s(\lambda; \bw)\, \int \mathcal{D}_n(\psi, \psibar)
 \, \frac{(\psibar, \psi)^{s}  }{s!} \nonumber \\
& \,\exp\left[{n \sum_{k\ge 2} w_k \,\frac{ (\psibar, \psi)^{k-1} }{(k-1)!} - (\psibar, J \psi) \sum_{k\ge 2} w_k \,\frac{ (\psibar, \psi)^{k-2} }{(k-2)!}}\right] \\
=   \sum_{s\ge 1}\, & \Pi_s(\lambda; \bw) \, t_{n,s}(\bw) \\
=   \sum_{s\ge 1}\, & {n-1 \choose s-1}\, \Pi_s(\lambda; \bw)\,  P_{n-s}(n; \bw)
\end{align}
The total weight of unrooted hyperforests on the set on $n$ vertices, irrespective from the number of hypertrees, is obtained from the partition function at $\lambda=1$
\be
F_n(\bw) := F_n(1; \bw)  =  \sum_{s\ge 1}\, \Pi_s(1; \bw) \, t_{n,s}(\bw)\, .  \label{def:F}
\ee
Also we get
\begin{align}
u_{n,p} = u_{n,p} (\bw) = & \sum_{s\ge 1}\, \pi_{s,p}(\bw) \, t_{n,s}(\bw) \label{eq:unp}\\
 = &
\sum_{s\ge 1}\,  {n-1 \choose s-1}\, \pi_{s,p}(\bw)\,  P_{n-s}(n; \bw)
\end{align}

Remark  that from the definition 
\be
\pi_{s,p}(\bw) = 0 \qquad \hbox{ when } p > s
\ee
so that $\Pi_s(\lambda; \bw)$ is a polynomial of degree $s$. 
It is monic because 
\be
\pi_{s,s}(\bw)=1\ef .
\ee
And remark also that $\pi_{s,0}(\bw)=0$ while 
\be
\pi_{s,1}(\bw) = \begin{cases}
1 & \hbox{ for } s=1 \\
w_s \,(1-s) & \hbox{ otherwise.}
\end{cases}
\ee
Accordingly $u_{n,0}(\bw)=0$ and $u_{n,n}(\bw)=1$, while
it follows that the weight of unrooted hypertrees on $n$ vertices is simply the weigth of the rooted hypertrees divided by $n$, indeed from \reff{eq:unp} 
\begin{align}
u_n(\bw) := u_{n,1}(\bw) = &\, P_{n-1}(n; \bw) + \sum_{s\ge 2} w_s \,(1-s)\,{n-1 \choose s-1 } \, P_{n-s}(n; \bw) \\
= & \, P_{n-1}(n; \bw) - \, (n-1)\, \sum_{s\ge 2} w_s \,{n-2 \choose s-2 } \, P_{n-s}(n; \bw) \\
= & \, \frac{P_{n-1}(n; \bw)}{n} \\
= & \, \frac{t_n(\bw)}{n}\label{un1}
\end{align}
where we used the recursion relation \reff{id:P} for the polynomials $P_s(x; \bw)$ at $x=n$ and $s=n-1$.

More formally we can follow a different strategy. Let $D = \frac{\partial}{\partial t}$ then
\be
\exp \left[ \lambda \left(  y  + \sum_{k\ge 2} w_k\, (1-k)\,\frac{y^{k}}{k!} \right) \right] \, = \,
\left. \exp\left[\lambda\,  \sum_{k\ge 2} w_k \,(1-k)\frac{ D^{k} }{k!}\right] \,\exp\left(t\, y\right)\right|_{t=\lambda}
\ee
so that 
\be
\Pi_s(\lambda, \bw) = \left. \exp\left[\lambda\,  \sum_{k\ge 2} w_k \,(1-k)\frac{ D^{k} }{k!}\right] \,t^s\,\right|_{t=\lambda}
\ee
and therefore
\begin{align}
 F_n& (\lambda; \bw)   =  \nonumber\\
= & \exp\left[\lambda\,  \sum_{k\ge 2} w_k \,(1-k)\frac{ D^{k} }{k!} + n \sum_{k\ge 2} w_k \,\frac{ D^{k-1} }{(k-1)!}\right] 
\, \left[1 - \,\sum_{k\ge 2} w_k \, \frac{ D^{k-1} }{(k-2)!}\right] \nonumber\\
& \qquad  \int \mathcal{D}_n(\psi, \psibar) \, \left. e^{t \,(\psibar, \psi)} \right|_{t=\lambda}\label{fe1} \\
 = &  \exp\left[\lambda\,  \sum_{k\ge 2} w_k \,(1-k)\frac{ D^{k} }{k!} + n \sum_{k\ge 2} w_k \,\frac{ D^{k-1} }{(k-1)!}\right] 
\, \left.\left[ t^n - \,n\, \sum_{k\ge 2} w_k \, {n-1 \choose k-2}\, t^{n-k+1}\right] \right|_{t=\lambda} \nonumber
\end{align}
now,  we expand first the second exponential,  to get once more
\begin{align}
 F_n (\lambda; \bw)   
 = &  \left.  \exp\left[\lambda\,  \sum_{k\ge 2} w_k \,(1-k)\frac{ D^{k} }{k!} \right] \, E_n(t, \bw)\right|_{t=\lambda}\label{fe} \\
  = &   \exp\left[\lambda\,  \sum_{k\ge 2} w_k \,(1-k)\frac{ D^{k} }{k!} \right] 
\, \left.\left[ \sum_{s\ge 0}\,{n-1\choose s-1} \,P_{n-s}(n, \bw)\,t^s\right] \right|_{t=\lambda}\nonumber \\
 = &  \sum_{s\ge 0}\,{n-1\choose s-1}\, \Pi_s(\lambda, \bw)\,\,P_{n-s}(n, \bw)\nonumber \ef .
 \end{align}
\subsection{On the $k$-uniform complete hypergraph}   \label{sec:hyperforests}

When $\bw = \ve_k$ the formula \reff{def:Pi} becomes
 \be
\exp \left[ \lambda \left(  y  + \, (1-k)\,\frac{y^{k}}{k!} \right) \right] \, = \,
\sum_{s\ge 0} \, \Pi_s(\lambda; \ve_k) \,\frac{y^s}{s!} \, = \, \sum_{s\ge 0} \,  \sum_{p\ge 0}\, \pi_{s,p}(\ve_k) \,\lambda^p\, \frac{y^s}{s!} \ef .
\ee
We introduce a family of generalized Hermite polynomials $H_s^{(k)}(x)$ as defined by the generating function
\be
\exp \left[ x\, z  + \, (1-k)\,\frac{z^{k}}{k!}  \right] \, = \,\sum_{s\ge 0} \, H_s^{(k)}(x) \,\frac{z^s}{s!} \label{def:H}
\ee 
which when $k=2$ are related to the ordinary Hermite polynomials $H_s$ by
\be
H_s^{(2)}(x) = \He_s(x) = \frac{1}{2^{\frac{s}{2}}}\,H_s\left(\frac{x}{2^{\frac{1}{2}}}\right)\, . 
\ee
where $\He_s$ are sometimes used~\cite{AS}.
Similar generalizations of the Hermite polynomials can be found in~\cite{sub,Dj1,Dj2}. 
We then get
\be
\Pi_s(\lambda; \ve_k)  = \lambda^{\frac{s}{k}} \, H_s^{(k)}\left( \lambda^{\frac{k-1}{k}} \right) \label{eq:telP}\ef .
\ee
Thus the generating function of unrooted hyperforests is 
\be
F_n(\lambda; \ve_k) = \sum_{\substack{ p\ge 0 \\ p: \, (n-p) | (k-1) }}  {n -1 \choose p -1 }\, \frac{(n-p)!}{ \left(\frac{n-p}{k-1}\right)! [(k-1)!]^\frac{n-p}{k-1} } \, n^{\frac{n-p}{k-1}}\, \lambda^{\frac{p}{k}}\,  H_p^{(k)} \left( \lambda^{\frac{k-1}{k}} \right) \label{wuf}
\ee
where the sum is restricted to the values of $p$ such that  $n-p$ can be divided by $k-1$.
By using~\reff{fe1} we get instead
 \begin{align*}
F_n (\lambda; \ve_k)  
= &  \exp\left[\lambda \,(1-k)\frac{ D^{k} }{k!} + n  \,\frac{ D^{k-1} }{(k-1)!}\right] \, \left.\left[ t^n - \,n\,  {n-1 \choose k-2}\, t^{n-k+1}\right] \right|_{t=\lambda} \nonumber \\
= & \lambda^\frac{n}{k}\, \exp\left[ \,(1-k)\frac{ D^{k} }{k!} + \frac{n}{\lambda^\frac{k-1}{k} }  \,\frac{ D^{k-1} }{(k-1)!}\right] \, \left.\left[ t^n - \,\frac{n}{\lambda^\frac{k-1}{k} }\,  {n-1 \choose k-2}\, t^{n-k+1}\right] \right|_{t=\lambda^\frac{k-1}{k} } \nonumber \\
= & \lambda^\frac{n}{k}\, \exp\left[ \, \frac{n}{\lambda^\frac{k-1}{k} }  \,\frac{ D^{k-1} }{(k-1)!}\right] \, \left.\left[ H_n^{k}(t) - \,\frac{n}{\lambda^\frac{k-1}{k} }\,  {n-1 \choose k-2}\, H_{n-k+1}^{k}(t) \right] \right|_{t=\lambda^\frac{k-1}{k} } \nonumber \ef .
 \end{align*}
 In the particular case $k=2$ we soon get
 \be
F_n (\lambda; \ve_2) =   \, \sqrt{\lambda}^n\,\left[ \, \He_n\left(\sqrt{\lambda}+\frac{n}{\sqrt{\lambda}}\right) - \,\frac{n}{\sqrt{\lambda}}\,\He_{n-1}\left(\sqrt{\lambda}+\frac{n}{\sqrt{\lambda}}\right)\right]
\ee
 because $ \exp\left[ \alpha  \,\frac{\partial}{\partial t} \right] $ is the translation operator from $t$ to $t+\alpha$. The same result can be obtained by using~\reff{fe} and~\reff{e2} as
 \begin{align*}
F_n (\lambda; \ve_2)  
= &  \left.  \exp\left[- \lambda\, \frac{ D^{2} }{2} \right] \, E_n(t, \ve_2)\right|_{t=\lambda}\\ 
= &  \left.  \exp\left[- \lambda\, \frac{ D^{2} }{2} \right] \, \left[ (t+n)^n - n\,(t+n)^{n-1}\right]\right|_{t=\lambda}\\
= &  \left.  \exp\left[- \, \frac{ D^{2} }{2} \right] \, \sqrt{\lambda}^n\,
\left[\left(t+\frac{n}{\sqrt{\lambda}}\right)^n - \frac{n}{\sqrt{\lambda}}\,\left(t+\frac{n}{\sqrt{\lambda}}\right)^{n-1}\right]\right|_{t=\sqrt{\lambda}}\\
= & \, \sqrt{\lambda}^n\,\left[ \, \He_n\left(\sqrt{\lambda}+\frac{n}{\sqrt{\lambda}}\right) - \,\frac{n}{\sqrt{\lambda}}\,\He_{n-1}\left(\sqrt{\lambda}+\frac{n}{\sqrt{\lambda}}\right)\right]\, .
 \end{align*} 
This formula has been reported in~\cite{Takacs} for $\lambda=1$,  where it counts the total number of unrooted forests. In this case~\reff{wuf} becomes instead
\be
F_n(\ve_2) = \sum_{p\ge 1} {n -1 \choose p -1 }\,n^{n-p}\, \He_p(1)
\ee
in agreement with what obtained in~\cite{Takacs} and reported as the series A001858 in the {\em The On-Line Encyclopedia of Integer Sequences} by Sloane~\cite{Sloane}.

By using $D = \frac{\partial}{\partial x}$ we get 
\be
\exp \left[ x\, z  + \, (1-k)\,\frac{z^{k}}{k!}  \right] \, = \, \exp  \left[ \frac{1-k}{k!} \, D^k \right] \, \exp \left[ x\, z\right]
\ee 
and therefore
\begin{eqnarray}
H_s^{(k)}(x)  & = & \exp  \left[ \frac{1-k}{k!} \, D^k \right] \, x^s \\
& = & \sum_{q\ge 0} \frac{1}{q!} \left( \frac{1-k}{k!} \right)^q D^{kq} \, x^s \\
& = & \sum_{q\ge 0} \frac{1}{q!} \left( \frac{1-k}{k!} \right)^q \, \frac{s!}{(s - kq)!}\, x^{s - kq}
\end{eqnarray}
which implies because of \reff{eq:telP}
\be
\Pi_s(\lambda; \ve_k)  = \sum_{q\ge 0} \frac{1}{q!} \left( \frac{1-k}{k!} \right)^q \, \frac{s!}{(s - kq)!}\, \lambda^{s - (k-1)q}
\ee
so that
\be
\pi_{s,p}(\ve_k) = \sum_{q\ge 0} \frac{1}{q!} \left( \frac{1-k}{k!} \right)^q \, \frac{s!}{(s - kq)!}\, \delta_{p,s -  (k-1)q}
\ee
and therefore, by using \reff{eq:unp} 
\begin{eqnarray}
 u_{n,p}(\ve_k) &=&   \sum_{q\ge 0} t_{n,p+q(k-1)}(\ve_k)\, \frac{\left[p+q(k-1)\right]!}{(p-q)!}\, \frac{1}{q!} \left( \frac{1-k}{k!} \right)^q \\
&=& \frac{(n-1)!}{p!} \left[\frac{n}{(k-1)!}\right]^\frac{n-p}{k-1} \sum_{q=0}^p {p\choose q} \frac{p+(k-1)q}{\left(\frac{n-p}{k-1} -q\right)!} \left(\frac{1-k}{k n}\right)^q
\end{eqnarray}
when $n-p$ can be divided by $k-1$, otherwise it vanishes,
where we used the relation \reff{eq:t} and the explicit expression \reff{pris}. 
Once more in the simpler case $k=2$ this formula reduces to
\be
u_{n,p}(\ve_2) = \frac{1}{p!}  \sum_{q=0}^p \left(-\frac{1}{2}\right)^q  {p\choose q} {n-1 \choose p+q-1}\, n^{n-p-q}\, (p+q)!
\ee
a result which can be found in~\cite{Bollobas,results}.

In order to proceed we need the sums
\begin{align}
\frac{1}{p!}\, \sum_{q=0}^p {p\choose q} \frac{(-z)^{-q}}{(v-q)!} = & \frac{(-z)^{-p}}{v!} \, L_p^{(v-p)}(z)\\
\frac{1}{p!}\, \sum_{q=0}^p {p\choose q} \frac{q\,(-z)^{-q}}{(v-q)!} = & - z \, \frac{d}{dz}\,\frac{(-z)^{-p}}{v!} \, L_p^{(v-p)}(z)\\
= & \frac{(-z)^{-p}}{v!} \,\left[ p\, L_p^{(v-p)}(z) + z\, L_{p-1}^{(v-p+1)}(z) \right]\\
= &  \frac{(-z)^{-p}}{v!} \, v\, L_{p-1}^{(v-p)}(z)
 \end{align}
where $L_m^{(\alpha)}(x)$ are the associated Laguerre polynomials 
\be
 L_m^{(\alpha)}(x) := \sum_{\nu=0}^\infty {m+\alpha \choose m-\nu} \frac{(-x)^\nu}{\nu!}
 \label{def_L}
 \ee
which satisfy the recursion relation
\be
L_{p-1}^{(k)}(z) = \frac{1}{z}\, \left[p \, L_{p}^{(k)}(z) - (p+k)\, L_{p-1}^{(k)}(z)\right]
\ef .
\ee

We arrive at the representation
\begin{eqnarray}
u_{n,p}(\ve_k) & = &  \frac{(n-1)!}{ \left(\frac{n-p}{k-1}\right)!} \left[\frac{n}{(k-1)!}\right]^\frac{n-p}{k-1}\, \left(-\frac{k-1}{k\, n}\right)^p \nonumber \\
& &  \qquad \left[ p\, L_p^{\left(\frac{n-p}{k-1}-p\right)}\left(\frac{k\, n}{k-1}\right) + (n-p) \,
L_{p-1}^{\left(\frac{n-p}{k-1}-p\right)}\left(\frac{k\, n}{k-1}\right)\right]
\end{eqnarray}
for the number of unrooted hyperforests with $p$ hypertrees on the $k$-uniform complete hypergraph ${\cal K}_n^{(k)}$ with $n$ vertices. 

In order to study the asymptotic behaviour of the previous expression in the limit of large $n$ at fixed $p$ we need the following expansion for the Laguerre polynomial 
\be
L_s^{\left(\frac{n-p}{k-1}-p\right)}\left(\frac{k\, n}{k-1}\right) \simeq \frac{(-n)^s}{s!} \left\{ 1 + \frac{s\,[s+1 + 2\,k\,(p-s)]}{2\,n\,(k-1)} +O\left(\frac{1}{n^2}\right) \right\} \label{Lagu}
\ee
that can be easily obtained from the definition~\reff{def_L}, as shown in Appendix~\ref{app:lag}, then
\be
p\, L_p^{\left(\frac{n-p}{k-1}-p\right)}\left(\frac{k\, n}{k-1}\right) + (n-p) \,
L_{p-1}^{\left(\frac{n-p}{k-1}-p\right)}\left(\frac{k\, n}{k-1}\right) \simeq \frac{(-n)^p}{(p-1)!}\,\frac{1}{n}\, \frac{k}{k-1}
\ee
because the leading terms in the two contributions cancel out. We get
\be
u_{n,p}(\ve_k) \simeq { n-1\choose p-1} \, \frac{(n-p)!}{\left(\frac{n-p}{k-1}\right)!}\, 
\frac{
n^{\frac{n-p}{k-1}-1} }{\left[(k-1)!\right]^{\frac{n-p}{k-1}}} \,\left(\frac{k-1}{k}\right)^{p-1} \label{ask}
\ee
Remark that when $p=1$ this formula is exact, indeed
\be
u_{n}(\ve_k) = u_{n,1}(\ve_k) = \frac{(n-1)!}{\left(\frac{n-1}{k-1}\right)!}\, 
\frac{
n^{\frac{n-1}{k-1}-1} }{\left[(k-1)!\right]^{\frac{n-1}{k-1}}} = \frac{t_{n,1}(\ve_k)}{n}
\ee
is the number of unrooted hypertrees in $n$ vertices, because of the general result~\reff{un1} and the explicit expression~\reff{pris}. In~\cite{Karonski_02} this number is quoted as obtained in~\cite{Selivanov}.

The formula~\reff{ask} at $k=2$  provides the result
\be
u_{n,p}(\ve_2) \simeq {n-1 \choose p-1} \frac{n^{n-p-1}}{2^{p-1}}
\ee
already obtained in~\cite{results} by a different method. It follows that the partition function is, if $\lambda$ is such that the relevant contribution to the sum comes from regions which don't change with $n$,  a problem which we will discuss elsewhere, we get
\begin{eqnarray*}
\sum_{p=0}^\infty u_{n,p}(\ve_2) \,  \lambda^p &\sim& n^{n-2}\, \lambda\,\sum_{p=0}^{n-1} {n-1 \choose p} \left( \frac{\lambda}{2\,n}\right)^p = n^{n-2}\, \lambda\, \left(1 +  \frac{\lambda}{2\,n}\right)^{n-1}\\
& \simeq & n^{n-2}\, \lambda\,e^{\frac{\lambda}{2}}
\end{eqnarray*}
which at $\lambda=1$ provides the well-known result by~\cite{Renyi, Denes}.

More generally, by using the Stirling approximation for large factorials
\be
u_{n,p}(\ve_k)   \simeq  \frac{n^{n-2}}{e^{n\frac{k-2}{k-1}}}
\frac{\sqrt{k-1}}{\left[(k-2)!\right]^\frac{n-p}{k-1}}
 \frac{1}{(p-1)!}  \,\left(\frac{k-1}{k}\right)^{p-1}
\ee
while
\be
\sum_{p=0}^\infty u_{n,p}(\ve_k) \, \lambda^p \simeq  \frac{n^{n-2}}{e^{n\frac{k-2}{k-1}}}
\frac{\sqrt{k-1}}{\left[(k-2)!\right]^\frac{n}{k-1}}
 \, \lambda\,e^{\frac{k-1}{k} [(k-2)!]^{\frac{1}{k-1}}\lambda}
\ee

\subsection{On the complete hypergraph}\label{final}

When $\bw = \bone$ \reff{def:Pi} becomes
 \be
\exp\left[\lambda\, ( 1-y ) \, \left(e^y-1\right)\right]   \, = \, 
\sum_{s\ge 0} \, \Pi_s(\lambda; \bone) \,\frac{y^s}{s!} \, = \, \sum_{s\ge 0} \,  \sum_{p\ge 0}\, \pi_{s,p}(\bone) \,\lambda^p\, \frac{y^s}{s!} 
\ee
%
Now
\begin{eqnarray}
\pi_{s,p} ( \bone) & = & s!\, [y^s] [\lambda^p] \,\exp\left[\lambda\, ( 1-y ) \, \left(e^y-1\right)\right]   \\
& = &  s!\, [y^{s}] \, ( 1-y )^p \,\frac{ \left(e^y-1\right)^p}{p!} \\
& = &  s!\, [y^{s}] \,  \sum_{m\ge 0}\, (-1)^m\, {p \choose m} \,y^m\,  \sum_{q\ge 0} \left\{ {q + p \atop p} \right\} \frac{y^{q+p}}{(q+p)!}  \\
& = & \sum_{q\ge 0}\, (-1)^{s-p-q}\, {p\choose s-p-q} \, \left\{ {p+q \atop p} \right\} \frac{s!}{(p+q)!} 
\end{eqnarray}
so that the number of unrooted hyperforests with $p$ hypertrees obtained by formula  \reff{eq:unp}, by using the number of  rooted hyperforests given in \reff{tbone}, is 
\begin{align}
u_{n,p} &(\bone)   =   \\
= & \sum_{s \ge 1} {n-1 \choose s-1} \,  b_{n-s}(n) \sum_{q\ge 0}\, (-1)^{s-p-q}\, {p\choose s-p-q} \, \left\{ {p+q \atop p} \right\} \frac{s!}{(p+q)!} \nonumber \\
= & \sum_{s \ge 1} {n-1 \choose s-1} \,  \sum_{r\ge 0}\, \left\{ {n-s \atop r} \right\} \,n^r\, \sum_{q\ge 0}\, (-1)^{s-p-q}\, {p\choose s-p-q} \, \left\{ {p+q \atop p} \right\} \frac{s!}{(p+q)!} \nonumber\ef .
\end{align}
Of course, because of the general result~\reff{un1},
\be
u_{n} (\bone) = u_{n,1} (\bone)   = \frac{t_{n,1} (\bone) }{n} =  \frac{b_{n-1} (n) }{n} = \sum_{r\ge 0} \, \left\{ {n-1 \atop r} \right\}\, n^{r-1}
\ee
a sequence which is reported with the number A030019 in the {\em The On-Line Encyclopedia of Integer Sequences} by Sloane~\cite{Sloane}.

\section{Conclusions}\label{conclusions}

We have studied the generating function of both rooted and unrooted hyperforests in the complete hypergraph with $n$ vertices, when the weight of each hyperedge depends only on its cardinality.
All the results could also be obtained by starting from recursion relations in the number of vertices,  to obtain implicit relations for the formal power series of the generating function,
which can be afterwards solved by using the Lagrange inversion formula. However we showed here how the same problem can be directly and more easily solved by means of a novel Grassmann representation.

Once we obtained the general solutions we have restricted to particular cases to recover more explicit results. In particular we considered the case of the $k$-uniform complete hypergraph, where only edges of cardinality $k$ are present. When this weight is set to one we are reduced to a counting problem. We thus obtained a generalization of many known results in the case $k=2$ namely of ordinary forests on the complete graph. In the case of unrooted hyperforests we also recovered a novel explicit expression for their number with $p$ connected components, that is hypertrees, in terms of the associated Laguerre polynomials, for any $k$. We have also presented the asymptotic behaviour of these numbers for large number of vertices.

A second direct application of the general solutions is obtained for the complete hypergraph when all the hyperedges have the same weights.

\section*{Acknowledgements}

It is a pleasure to thank Alan~D.~Sokal for his interest and his suggestions on how to improve our results and our presentation.

\appendix
 
 \section{Stirling and Bell numbers, Bell polynomials}\label{app:num}
 
The Stirling numbers of the second kind, denoted by $\left\{ {n \atop k} \right\}$  according to the notation  introduced in 1935 Jovan Karamata and promoted later by Donald Knuth, 
stands for the number of ways to partition a set of cardinality $n$ into $k$ {\em nonempty} subsets. Thus
\begin{eqnarray*}
\left\{ {n \atop k} \right\}  = 0 & & \qquad \hbox{ for } n < k\\
\left\{ {n \atop 0} \right\}  = 0 & & \qquad \hbox{ for } n \ge 1\\
\left\{ {0 \atop 0} \right\}  = 1 & & \\
\end{eqnarray*}
if we agree that there is one way to partition an empty set into zero nonempty parts. 
Chosen an object among  $n>0$  to be partitioned into $k$ nonempty parts, we either put it into a class by itself (in $\left\{ {n-1 \atop k-1} \right\}$ ways) or we put it together with some nonempty subset of the other $n-1$ objects (there are $k \left\{ {n-1 \atop k} \right\}$ possibilities, because each of the $\left\{ {n-1 \atop k} \right\}$ ways to partition the $n-1$ other objects into $k$ nonempty parts gives $k$ subset that the chosen object can join), hence we get the recurrence
\be
\left\{ {n \atop k} \right\} = k\,\left\{ {n-1 \atop k} \right\} + \left\{ {n-1 \atop k-1} \right\} 
\label{rec:s}
\ee
which enables us to compute them.

Their exponential generating function is
\be
\sum_{n\ge 0}\left\{ {n \atop k} \right\}  \frac{z^n}{n!} = \sum_{n\ge k}\left\{ {n \atop k} \right\}  \frac{z^n}{n!} = \frac{\left(e^z -1\right)^k}{k!}
\ee

 The Bell number $b_n$ is the number of all possible subsets of a set of cardinality $n$, hence
 \be
 b_n = \sum_{k\ge 0} \left\{ {n \atop k} \right\} \, . 
 \ee
 Their exponential generating function is
\be
\sum_{n\ge 0} b_n \frac{z^n}{n!} = \sum_{n\ge 0}  \sum_{k\ge 0} \left\{ {n \atop k} \right\} \frac{z^n}{n!} =  \sum_{k\ge 0} \sum_{n\ge 0}  \left\{ {n \atop k} \right\} \frac{z^n}{n!} =
 \sum_{k\ge 0} \frac{\left(e^z -1\right)^k}{k!} = e^{e^z -1}\label{rec:b}
\ee

The Bell polynomials, also called exponential polynomials, are given by
\be
b_n(x) := \sum_{k\ge 0} \left\{ {n \atop k} \right\} x^k
\ee
so that
\be
b_n(1) = b_n
\ee
and they satisfy the recurrence relation
\be
b_{n+1}(x) = x \left[ b_n(x) +  b_n'(x)\right]\label{rec:bn}
\ee
as follows from~(\ref{rec:s}).

Their exponential generating function is
\be
\sum_{n\ge 0} b_n(x) \frac{z^n}{n!} = \sum_{n\ge 0}  \sum_{k\ge 0} \left\{ {n \atop k} \right\} x^k \frac{z^n}{n!} =  \sum_{k\ge 0} \frac{ \left[ x\left(e^z -1\right)\right]^k}{k!} = e^{x (e^z -1)} \label{egf:bell}
\ee


 \section{Exponential generating function}\label{app:egf}
 
 Counting the number of unrooted trees $u_n$ on the complete graph ${\cal K}_n^{(2)}$ is presented in~\cite[Chapter 7]{Graham_94} as a simple application of the formalism of the exponential generating function.
 
 For $n>0$ the recurrence
 \be
 u_n = \sum_{m>0}\,\frac{1}{m!} \sum_{\substack{a_1,a_2, \cdots, a_m \\ a_1+\cdots +a_m=n-1}} {n-1 \choose a_1, \cdots, a_m}\, a_1\cdots  a_m\, u_{a_1}\cdots u_{a_m}
 \label{eq:trec}
 \ee
can be obtained as follows.
A given vertex is attached to $m$ components of sizes $a_1, \cdots, a_m$. There are
${n-1 \choose a_1, \cdots, a_m}$ ways to assign $n-1$ vertices to those components and $ a_1\cdots a_m$ ways to connect the given vertex to them. There are $u_{a_1}\cdots u_{a_m}$ ways to connect those individual components with spanning trees; and we divide by $m!$ because the $m$ components are not ordered. 

As the number of rooted trees is
\be
t_n = n\, u_n
\ee
the recurrence relation can be re-written as
\be
\frac{t_n}{n!} = \sum_{m>0} \,\frac{1}{m!} \sum_{\substack{a_1,a_2, \cdots, a_m \\ a_1+\cdots +a_m=n-1}} \frac{t_{a_1}}{a_1!}\cdots \frac{t_{a_m}}{a_m!}\, .
\label{eq:srec}
 \ee
By introducing the exponential generating function for the sequence $\{t_n\}$
\be
T(z) := \sum_{n\ge 0} t_n\, \frac{z^n}{n!}
\ee
it follows that the inner sum in~(\ref{eq:srec}) is the coefficient of $z^{n-1}$ in $T(z)^m$
\be
\frac{t_n}{n!} = \,\left[z^{n-1}\right] \,\sum_{m\ge 0} \frac{1}{m!}\, T(z)^m = 
 \,\left[z^{n-1}\right] \,e^{T(z)}
\ee
where we have included also the case $n=1$ by adding the contribution $m=0$.
And therefore
\be
T(z) = z\,e^{T(z)}
\ee
so that because of~\reff{expt} and~\reff{propexpt} $T$ is related to the generalized exponential $\cal E$ by
\be
T(z) = z\, {\cal E}(z)\, .
\ee
$T$ is also related to the Lambert $W$ function~\cite{CJK} by
\be
T(z) = - W(-z)\, .
\ee
 Now
 \be
 u_n = \frac{t_n}{n} = \frac{n!}{n}\, \left[z^{n-1}\right] \, {\cal E}(z) = n^{n-2}\, .
 \ee
 This result is usually attributed to Cayley in 1889~\cite{Cayley}, but in his paper he refers to a previous result by Borchardt in 1860~\cite{Borchardt}.
 
 More generally when $\theta(u)$ is a formal power series in $u$ with $\theta(0)=1$, a relation for the formal power series $T(z)$ of the form
 \be
 T(z) = z \, \theta \left( T(z) \right)
 \ee
 has a unique solution, which is given by Lagrange inversion formula~\cite{Wilf}
 \be
 \left[z^n\right] T(z) = \frac{1}{n}  \left[T^{n-1}\right] \theta\left( T \right)^n\, .
 \ee
 Furthermore
 \be
 \left[z^n\right] T(z)^r = \frac{r}{n}  \left[T^{n-r}\right] \theta\left( T \right)^n\, .
 \ee

 In our application to the trees,  $\theta(T) = e^T$ and therefore
 \be
 u_n =  \frac{t_n}{n} = \frac{n!}{n}\,  \left[z^n\right] T(z) = \frac{(n-1)!}{n}\,   \left[T^{n-1}\right] e^{n \, T} = \frac{n^{n-1}}{n}\, .
 \ee
While the number of rooted forests with $r$ trees is given by
 \be
t_{n,r} = \frac{n!}{r!}\, \left[z^n\right] T(z)^r =  \frac{(n-1)!}{(r-1)!}\,  \left[T^{n-r}\right] e^{n \, T} = {n-1 \choose r-1}\,n^{n-r}\, .
 \ee

More generally, in the case of the $k$-uniform complete hypergraph ${\cal K}_n^{(k)}$ with weights $w_k$ the recurrence relation for the weight of unrooted hypertrees is 
 \be
u_n = \sum_{\substack{m>0\\ m / (k-1)}}\,\frac{w_k^\frac{m}{k-1}}{\left(\frac{m}{k-1}\right)!\, \left[(k-1)!\right]^\frac{m}{k-1}}\, \sum_{\substack{a_1,a_2, \cdots, a_m \\ a_1+\cdots +a_m=n-1}} {n-1 \choose a_1, \cdots, a_m}\, a_1\cdots a_m\, u_{a_1}\cdots u_{a_m}
 \label{eq:tdrec}
 \ee
where at variance with respect to~(\ref{eq:trec}) the sum on $m$ is restricted to integers that can be divided by $k-1$ and appears a combinatorial factor  $\frac{m!}{\left(\frac{m}{k-1}\right)!\, \left[(k-1)!\right]^\frac{m}{k-1}}$ because this is the number of ways in which the $m$ sub-hypertrees can be hooked to the starting vertex by using hyperedges of cardinality $k$. As a consequence the equation for the rooted hypertrees becomes
\be
\frac{t_n}{n!} = \,\left[z^{n-1}\right] \,\sum_{l \ge 1} \frac{w_k^l}{\left[(k-1)!\right]^{l}\, l!}T(z)^{(k-1)\,l} =
 \,\left[z^{n-1}\right] \,e^{w_k\,\frac{T(z)^{k-1}}{(k-1)!}}
\ee
which is to say
\be
T(z) = \,z\,e^{w_k\,\frac{T(z)^{k-1}}{(k-1)!}}
\ee
We can now apply the Lagrange inversion formula with  $\theta(T) = e^{w_k\,\frac{T(z)^{k-1}}{(k-1)!}}$ and therefore
 \be
 u_n =  \frac{t_n}{n} = \frac{n!}{n}\,  \left[z^n\right] T(z) = \frac{(n-1)!}{n}\,   \left[T^{n-1}\right] e^{n\,w_k\,\frac{T(z)^{k-1}}{(k-1)!}} = \frac{1}{n}\,\frac{\left(n\,w_k\right)^\frac{n-1}{k-1} }{\left( \frac{n-1}{k-1} \right)!  \left[(k-1)!\right]^\frac{n-1}{k-1}}\, .
 \ee
While the weight for the rooted hyperforests with $r$ hypertrees is 
 \be
t_{n,r} = \frac{n!}{r!}\, \left[z^n\right] T(z)^r =  \frac{(n-1)!}{(r-1)!}\,  \left[T^{n-r}\right] e^{n \, \frac{T^{k-1}}{(k-1)!}} =  \frac{(n-1)!}{(r-1)!}\,\frac{1}{\left(\frac{n-r}{k-1}\right)!}\,
\frac{\left(n \, w_k\right)^{\frac{n-r}{k-1}}}{\left[(k-1)!\right]^{\frac{n-r}{k-1}}}
 \ee
when $(n-r)/(k-1)$ is an integer. It is indeed the total number of hyperedges.

In the general case of the complete hypergraph $\overline{\cal K}_n$ the recurrence relation for the total weight of unrooted hypertrees is more involved, but the possibilities of attaching hyperedges of different cardinality at the starting vertex are mutually avoiding and this makes the recursion affordable. It follows that the generating function satisfies the equation
\be
T(z) = \,z\,e^{\sum_{k\ge 2} w_k\,\frac{T(z)^{k-1}}{(k-1)!} } \label{eq:tt}
\ee
so that
\be
t_{n,r} =   \frac{(n-1)!}{(r-1)!}\,  \left[T^{n-r}\right] e^{n \,\sum_{k\ge 2} \, w_k\, \frac{T^{k-1}}{(k-1)!}}\, =
{n-1 \choose r-1} P_{n-r} (\bw)
\ee
where we introduced the polynomials $P_{n-r} (\bw)$ of the weights $w_k$'s defined in~\reff{teta}.

In the simpler case in which all the weights are equal to, say, $x$, the recurrence relation for the unrooted hypertrees is 
 \be
 u_n = \sum_{m\ge 0}\sum_{l\ge 0}\,\frac{1}{m!}\,\left\{ {m\atop l}\right\} \, x^l\,\sum_{\substack{a_1,a_2, \cdots, a_m \\ a_1+\cdots +a_m=n-1}} {n-1 \choose a_1, \cdots, a_m}\, a_1\cdots a_m\, u_{a_1}\cdots u_{a_m}
 \label{eq:hrec}
 \ee
where, at variance with respect to~(\ref{eq:trec}) there appears a factor $\left\{ {m\atop l}\right\}$ because this is the number of ways in which the $m$ sub-hypertrees can be hooked to the starting vertex by using $l$ generic hyperedges. As a consequence the equation for the rooted hypertrees becomes
\be
\frac{t_n}{n!} = \,\left[z^{n-1}\right] \,\sum_{m\ge 0} \sum_{l\ge 0}\, \frac{1}{m!}\,\left\{ {m\atop l}\right\}\, x^l\,T(z)^m =
 \,\left[z^{n-1}\right] \,e^{x\left(e^{T(z)}-1\right)}
\ee
which is to say
\be
T(z) = \,z\,e^{x\left(e^{T(z)}-1\right)}
\ee
that is \reff{eq:tt} for $w_k=x$ for all $k$, a relation that  in the case $x=1$  is reported in the Warme's Ph.~D.~Thesis~\cite{Warme} as due to W.~D.~Smith, but see also~\cite{Gessel_05}.
We can now apply the Lagrange inversion formula with  $\theta(T) = e^{x\left(e^T-1\right)}$ and therefore
 \be
 u_n =  \frac{t_n}{n} = \frac{n!}{n}\,  \left[z^n\right] T(z) = \frac{(n-1)!}{n}\,   \left[T^{n-1}\right] e^{n\,x\,\left(e^T-1\right)} = \frac{b_{n-1}(n\, x)}{n}\, . 
 \ee
 While the total weight of rooted hyperforests with $r$ hypertrees is
\be
t_{n,r} = \frac{n!}{r!}\, \left[z^n\right] T(z)^r =  \frac{(n-1)!}{(r-1)!}\,  \left[T^{n-r}\right] e^{n \, x\,\left(e^T -1\right)} =  {n-1 \choose r-1}\,b_{n-r}(n\,x)\, .
\ee

 \section{Asymptotic behaviour of associated Laguerre polynomials}\label{app:lag}
 
%
 In this appendix we will study the asymptotic behaviour of the associated  Laguerre polynomial $L_s^{\left(\frac{n-p}{d-1}-p\right)}\left(\frac{d\, n}{d-1}\right)$ for large $n$.
 
 We remark that for $\alpha, \nu, s$ all integers and $\alpha \gg 1$
 \begin{eqnarray*}
 \frac{(s+\alpha)!}{(\nu+\alpha)!} & = & (s+ \alpha) \cdots (\nu + 1 + \alpha) \\
 & \simeq & \alpha ^{s -\nu} + \alpha ^{s -\nu - 1} \, \left[ s + \cdots + (\nu + 1) \right] \\
 & = & \alpha ^{s -\nu} + \alpha ^{s -\nu - 1} \, \left[ \frac{s ( s+1)}{2} - \frac {\nu (\nu +1)}{2}\right]
 \end{eqnarray*}
 if
 $$ 
 \alpha = \frac{n-p}{d-1}-p
 $$
 we get, for $n \gg 1$ at first order in $1/n$
 $$
 \alpha ^{s -\nu}  \simeq \left(\frac{n}{d-1}\right) ^{s -\nu}\,\left[ 1 - (s-\nu) \,  \frac{p\, d}{n} \right]
 $$
 and 
 $$
  \frac{(s+\alpha)!}{(\nu+\alpha)!}  \simeq \left(\frac{n}{d-1}\right) ^{s -\nu}\,\left\{ 1 - (s-\nu) \,  \frac{p\, d}{n} + \frac{d-1}{n}\, \left[ \frac{s ( s+1)}{2} - \frac {\nu (\nu +1)}{2}\right]
 \right\}
 $$
so that
\begin{eqnarray*}
L_s^{\left(\alpha\right)}\left(\frac{d\, n}{d-1}\right) & = & 
\sum_{\nu =0}^s  \frac{(s+\alpha)!}{(\nu+\alpha)!}  \, 
\frac{1 } {\nu ! \, (s-\nu) !} \, \left( - \frac{d\, n}{d-1}\right)^\nu
 \\
 & \simeq &  \left(\frac{n}{d-1}\right)^s \, \frac{1}{s!}\, \sum_{\nu =0}^s {s \choose \nu} \,(-d)^\nu \,\left\{ 1 + \vphantom{ \frac{s ( s+1)}{2}}\right. \\
 & & \left.-  (s-\nu) \,  \frac{p\, d}{n} + \frac{d-1}{n}\, \left[ \frac{s ( s+1)}{2} - \frac {\nu (\nu +1)}{2}\right] \right\}
\end {eqnarray*}
Now
$$
\sum_{\nu=0}^s \, {s \choose \nu} \,(-d)^\nu = (1 - d)^s
$$
and by taking one and two derivatives with respect to $-d$ we get
\begin{eqnarray*}
\sum_{\nu=0}^s\, \nu\, {s \choose \nu} \,(-d)^{\nu-1} & = & s\,(1 - d)^{s-1}\\
\sum_{\nu=0}^s\, \nu\,(\nu-1)\, {s \choose \nu} \,(-d)^{\nu-2} & = & s\,(s-1)\,(1 - d)^{s-2}
\end {eqnarray*}
and therefore
\begin{eqnarray*}
\sum_{\nu=0}^s\, \nu\,(\nu+1)\, {s \choose \nu} \,(-d)^{\nu} & = &   s\,(s-1)\,(1 - d)^{s-2}\, d^2 - 2\,s\,(1 - d)^{s-1}\,d \\
& = &   s\,(s+1)\,(1 - d)^{s-2}\, d^2 - 2\,s\,(1 - d)^{s-2}\,d
\end{eqnarray*}
and we get
\begin{eqnarray*}
L_s^{\left(\alpha\right)}\left(\frac{d\, n}{d-1}\right) & \simeq & 
\left(\frac{n}{d-1}\right)^s \, \frac{1}{s!}\,\left\{ (1 - d)^{s} - \left[s\,(1-d)^s + s\,(1-d)^{s-1}\,d\right] \frac{p\, d}{n}\right. \\
& &  \hspace{-3cm}+\, \frac{d-1}{n}\,  \left. \left[ \frac{s ( s+1)}{2} (1-d)^s -  \frac{s ( s+1)}{2} \, (1-d)^{s-2}\,d^2 + s\,(1-d)^{s-2}\,d\right]\right\}\\
& = & \frac{(-n)^s}{s!} \left\{ 1 + \frac{s\,(p+1)\, d}{(d-1)\,n} + \frac{s\,(s+1)\,(1-2\,d)}{2\,(d-1)\,n} \right\}
\end{eqnarray*}
from which~(\ref{Lagu}) follows.

\end{document}